\documentclass[aps,prd,twocolumn,groupedaddress,nofootinbib,longbibliography]{revtex4-1}
\usepackage{amsfonts,amsmath,amssymb}
\usepackage{nicefrac}
\usepackage{graphicx}
\usepackage{color}
\usepackage{hyperref}
\usepackage[normalem]{ulem}

\def\beq{\begin{equation}}
\def\eeq{\end{equation}}
\def\bea{\begin{eqnarray}}
\def\eea{\end{eqnarray}}

\begin{document}

%\preprint{}

\title{A Near Horizon Extreme Binary Black Hole Geometry}


\author{Jacob Ciafre$^{1}$}, \email[]{jakeciafre@outlook.com}  \author{Maria J.~Rodriguez$^{1,2}$}
\email[]{maria.rodriguez@usu.edu\\maria.rodriguez@aei.mpg.de}
\affiliation{$^{1}$Department of Physics, Utah State University, 4415 Old Main Hill Road, UT 84322, USA }
\affiliation{$^{2}$ Max Planck for Gravitational Physics, Potsdam 14476, Germany}

\begin{abstract}
A new solution of four-dimensional vacuum General Relativity is presented. It describes the near horizon region of the extreme (maximally spinning) binary black hole system with two identical extreme Kerr black holes held in equilibrium by a massless strut. This is the first example of a non-supersymmetric, asymptotically flat near horizon extreme binary black hole geometry of two uncharged black holes. The black holes are co-rotating, and the solution is uniquely specified by the mass. The binary extreme system has finite entropy. The distance between the black holes is fixed, but there is a zero-distance limit where the objects collapse into one. This limiting geometry corresponds to the near horizon extreme Kerr (NHEK) black hole.
\end{abstract}

%\keywords{}

\maketitle


Several black holes have been observed to spin at nearly the speed of light (the theoretical upper limit for stationary black holes) \cite{Gou:2011nq}-\cite{McClintock:2006xd}. Although there is considerable uncertainty in how black holes attain nearly extremal spins, several mechanisms in Nature can be envisioned. These include the merging of binary black holes (BBHs) possibly spinning at nearly extremal spins, and prolonged disk accretion. Numerical simulations, that are key in developing the most precise predictions of {\it dynamical} BBHs mergers \cite{Rezzolla:2007xa}-\cite{Lovelace:2010ne} and black hole disk accretion \cite{Volonteri:2004cf}-\cite{Shapiro:2004ud}, have confirmed the feasibility of these mechanisms giving rise to nearly extremal spinning black holes.

From a theoretical perspective, a rotating Kerr black hole obeys cosmic censorship: any singularities must lay behind the black hole's event horizon. As the black hole spin increases, for a fixed mass, the event horizon area protecting the singularity decreases. The maximum allowed spin, which corresponds to the speed of light, is reached before exposing the singularity. These maximally rotating black holes are called extreme and are the central objects of study of this work.

At extremality, the black hole's temperature drops to zero. Interestingly, at this point the geometry close to the event horizon, the so called Near-Horizon Extreme Kerr geometry \cite{Bardeen:1999px}, displays a very special feature: the symmetry is enhanced to an $SL(2,R)$, conformal, symmetry and develops a warped version of $AdS_3$. NHEK is obtained through scalings from the extreme Kerr black hole, has finite size, namely, finite event horizon area, and retains all the relevant aspects of black holes: an event horizon and an ergosphere. The enhanced conformal symmetry in NHEK, which does not extend to the full Kerr geometry, motivated works such as the Kerr/CFT conjecture concerning the quantum structure of black holes \cite{Guica:2008mu} and studies on the dynamics of stars \cite{Hadar:2014dpa} and energy extraction \cite{Lupsasca:2014pfa} in this region.
 
One then wonders, could NHEK be formed from a merger of the near horizon geometries of extreme BBHs? And, in this case, is the conformal symmetry also present in the geometry giving rise to NHEK after merger? To answer these questions, in this paper, we propose to focus on the stationary - rather than dynamical - BBHs and find explicitly the Near-Horizon Extreme Kerr Binary Black Hole geometry, ``NHEK2". We will show that this new metric retains only partially the symmetry of NHEK, including the dilatations, and are ``tree"-like geometries: they are solutions with NHEK asymptotics, but as one moves inward the geometry branches into two smaller warped $AdS_3$ regions.

Luckily, {\it stationary} (non-dynamical) BBHs solutions of Einstein's equations of General Relativity in vacuum with two-Kerr black holes are also known analytically. These are exact asymptotically flat vacuum solutions with two spinning (neutral) Kerr black holes supported by a conical singularity along the line separating the black holes. These were first found in \cite{Kinnersley:1978pz} and constructed via a variety of solutions generating techniques in i.e. \cite{Manko:2009zz}-\cite{Manko:2011ts}. The black holes in these solutions can co-rotate or spin in opposite directions. They become, for certain range of the parameters, extremal zero-temperature stationary BBHs solutions \cite{Kramer}-\cite{Manko:2011ts}. Remarkably, as first observed in \cite{Herdeiro:2008kq,Costa:2009wj}, the extreme BBHs overtake the extremal Kerr bound. All of these geometries were originally written in Weyl coordinates, where many expressions simplify. In these coordinates, the metric takes the form
\begin{equation}\label{Weyl}
ds^2=-\frac{\hat{\rho}^2}{f}d\hat{t}^2+f (d\hat{\phi}+\omega \,d\hat{t})^2+e^{2\nu}(d\hat{\rho}^2+d\hat{z}^2)\,,
\end{equation}
where $f,\nu,\omega$ are functions only of the $(\hat{\rho},\hat{z})$-coordinates, and $\hat{t}\in(-\infty,\infty)$, $\hat{\rho}, \hat{z} \in(-\infty,\infty)$ and $\hat{\phi}\sim\hat{\phi}+2\pi \Delta\hat{\phi}$
Note that all stationary axi-symmetric black hole solutions can be written in Weyl coordinates (\ref{Weyl}), including the stationary BBHs central to our discussion, and also the Kerr and NHEK geometries. Our starting point to find the NHEK2 geometry are the co-rotating extreme double-Kerr black hole solutions in \cite{Manko:2011ts}. We develop an appropriate near-horizon limiting procedure, and apply it to the extreme BBHs geometry. As a result we obtain the new NHEK2 geometries, which becomes a distinct object with finite entropy. Our results may be of considerable interest in extending the Kerr/CFT conjecture to BBHs systems, entropy calculations, and gravitational signatures of near rapidly spinning binary black holes.

{\bf The solution.} 
The solution of extremal co-rotating identical BBHs \cite{Manko:2011ts} from which we derive the new near-horizon geometry of extreme BBHs, takes a simpler representation in Weyl coordinates. We therefore  perform the scaling computations in this frame. In this case, we find that the appropriate near-horizon limiting procedure for the extremal BBHs is
\begin{gather}\label{rescalings1}
\rho =  \hat{\rho}\, \lambda  \,,\qquad z =  \hat{z}\,\lambda \ ,\\ t=\frac{\hat{t}}{\lambda}\,,\qquad \phi=\hat{\phi}+\frac{\hat{t}}{2M \lambda}\,,\\
p= -\frac{1}{\sqrt{2}}+\frac{(3\sqrt{2}-2)}{4} \, \lambda \qquad \kappa =  M \, \lambda\,,
\end{gather}
 taking $\lambda \rightarrow 0$ and keeping $( \hat{t},  \hat{\rho},  \hat{z},  \hat{\phi})$ fixed. In \cite{Manko:2011ts}, $p$ and $\kappa$ are free parameters related to the definitions of positive mass and the separation between the black holes respectively.
As a result of this process, we find the near-horizon extreme black hole binary NHEK2 geometry in Weyl coordinates (\ref{Weyl}). This is defined by the equations
\begin{small}
\begin{equation}
\begin{gathered}
\label{NHEK2sol2}
f=-\frac{4 M^2\mu_0\,(\mu_0+ 2\sigma_0^2)}{ \mu_0\,(\mu_0+2\sigma_0^2-2\sigma_1+\pi_0)+\mu_1\, \pi_1+(1-y^2)\,\sigma_0\,\tau_0},\,\\
\omega=\frac{ \pi_0 \,\sigma_0+\pi_{1}\,\sigma_1-\mu_1-4 \sigma_0 \,\sigma_1-(1-y^2)\,\tau_0/2}{2M(\mu_0+2\, \sigma_0^2)}\,,\\
e^{2\nu}=\frac{ \mu_0\,(\mu_0+2\sigma_0^2-2\sigma_1+\pi_0)+\mu_1\, \pi_1+(1-y^2)\,\sigma_0\,\tau_0}{K_0^2\,(x^2-y^2)^4}\,,
\end{gathered}
\end{equation}
\end{small}
where the functions yield
\begin{gather}
\mu_0= -\frac{\hat{\rho}^2}{2 M^2}\,,\qquad \sigma_0= -\frac{x^2-y^2}{2}-\frac{x^2+y^2}{\sqrt{2}}\,,
\end{gather}
\begin{gather}
\pi_1=-2\sqrt{2}\,x^2-(1-\sqrt{2})(x^2-y^2)\,,
\end{gather}
\begin{gather}
\mu_1=-\frac{(3-\sqrt{2})}{2}(-1+x^2)^2+(1-\sqrt{2})(x^2-y^2)^2\,,
\end{gather}
\begin{widetext}
\begin{gather}
\sigma_1=\frac{(3-\sqrt{2})}{2}(x^2-y^2)+\frac{(1-\sqrt{2})(3-2\sqrt{2})}{2\sqrt{2}}(x^2+y^2)\,,\\
\pi_0= \left(\frac{-74+85\sqrt{2}}{14}\right)x^2-\sqrt{2} x(1+x^2)+\left(\left(\frac{145-108\sqrt{2}}{28}\right)-x\right)(x^2-y^2)\,,\\ 
\tau_0= \sqrt{2} x(x^2-1)-\left(\frac{99-51\sqrt{2}}{28}-(1-\sqrt{2}) x\right)(x^2-y^2)+\left(\frac{71-23\sqrt{2}}{28}\right)(1-y^2)\,,
\end{gather}
\end{widetext}

and
\begin{eqnarray}
x &=& \frac{\sqrt{\hat{\rho}^2 + (\hat{z}+M)^2}  + \sqrt{\hat{\rho}^2 + (\hat{z} - M)^2} }{2 M } \,, \\
y &=& \frac{\sqrt{\hat{\rho}^2 + (\hat{z} + M)^2}  - \sqrt{\hat{\rho}^2 + (\hat{z} - M)^2} }{2M } \, ,\nonumber
\end{eqnarray}
with
\begin{eqnarray}
\label{NHEK2sol3}
K_0=-(1+\sqrt{2})/2\,.
\end{eqnarray}
This geometry arises as a scaling limit after a coordinate transformation of the extremal BBHs of \cite{Manko:2011ts}. We have checked that it is Ricci flat and, thus, a solution to the vacuum Einstein
equations. This solution represents the near horizon geometry of two extreme Kerr black holes. With this procedure, the asymptotically flat Minkowski region from the original metric decouples and the throat becomes infinitely long while there is a splitting into two pieces that survives. 

{\bf The event horizons.} Our solution displays two event horizons located at 
\begin{equation}
\hat{\rho}_H=0\,,\qquad \hat{z}_H=\pm M\,.
\end{equation}
Besides the conical singularity (to be analyzed in more detail below) the metric is smooth. The NHEK2 geometry is not asymptotically flat; in fact it has very peculiar asymptotics as $r\rightarrow \infty$ it approaches the NHEK asymptotic geometry, for $\hat{\rho}=r\sin\theta$ and $\hat{z}=r\cos\theta$.

In addition to the two $U(1)$ symmetries - $\partial_{\hat{t}}$ and $\partial_{\hat{\phi}}$ symmetries - present in the double-Kerr BBHs metric, the NHEK2 metric is invariant under $\hat{\rho} \rightarrow c \, \hat{\rho}$, $\hat{z} \rightarrow c \, \hat{z}$ and $\hat{t}\rightarrow c\,\hat{t}$ while $M\rightarrow c\,M$. So, the geometry (\ref{Weyl}) with (\ref{NHEK2sol2})-(\ref{NHEK2sol3}) has the dilation symmetry of $AdS_2$. However, the NHEK2 metric seems not to have all of the symmetries of $AdS_2$; the metric lacks one of the killing vectors that would otherwise close in $SL(2,R)$. We conclude then that its isometry group is not $SL(2,R)\times U(1)$. These enhancement is observed only when the two black holes collapse into one, and form a single larger black hole. In the collapse of NHEK2 to NHEK, the distance between the black holes goes to zero and corresponds to the $M\rightarrow 0$ limit in NHEK2 as we describe here below.\\
Inspection of the NHEK2 solution shows that changing coordinates as
\begin{gather}
\hat{t}\rightarrow   (-2+\sqrt{2}) \, M^2 \,T\,,\nonumber\\
\hat{\rho}\rightarrow (R \pm M) \sin\Theta \,,\nonumber\\ 
\hat{z}\rightarrow (R \pm M) \cos\Theta \mp M \,,\\ 
  \phi \rightarrow (-2+\sqrt{2}) \,T+ \frac{(148-107 \sqrt{2})}{56}M \,{\Phi}\,,\nonumber
\end{gather}
the expansions close to each of the black holes - located now at $R=\pm M$ - gives rise to a metric of the form
\begin{eqnarray}\label{eq:local}
ds^2 \sim M^2 \,\Gamma(\Theta) [ -(R\pm M)^2 dT^2 + \frac{dR^2}{(R\pm M)^2}\,\nonumber \\
+d\Theta^2+ \Lambda(\Theta) \left(d\Phi+(R\pm M) \, dT\right)^2 ]
 \end{eqnarray}
with
\begin{small}
\begin{gather}
\Gamma(\Theta) =\frac{\left((22-15 \sqrt{2})(3+\cos2 \Theta)\pm (72-52\sqrt{2})\cos\Theta\right)}{8}\,,\nonumber \\
\Gamma(\Theta)\Lambda(\Theta) = \frac{16 (2-\sqrt{2})^2 \sin^2\Theta}{\left((6-\sqrt{2})(3+\cos2\Theta)\pm(8-12\sqrt{2}\cos\Theta)\right)}\nonumber \,,
\end{gather}
\end{small}
when the leading terms in each metric component are retained. Hence locally, in the vicinity of each black hole, we note that the slices of the geometry at fixed polar angle $\Theta$ correspond to warped $AdS_3$. Our newly discovered metric are ``tree"-like geometries: they are solutions with NHEK asymptotics, but as one moves inward the geometry branches into two smaller warped $AdS_3$ regions. For the same special constant value of $\Theta=\Theta_0$ where $ \Lambda(\Theta_0)=1$, the local metrics are that of $AdS_3$. Other tree-like geometries have been previously found for extreme charged BBHs \cite{Maldacena:1998uz}. Whether this local manifestation of two copies of $AdS_3$ in NHEK hints really to an underlying conformal symmetry of the extreme BBHs is yet a question that needs further investigation.

The event horizon area can be computed from the metric (\ref{eq:local}) close to each of the black holes via
\begin{equation}
A_i\equiv 2\pi \int_{0}^{\pi} \sqrt{g_{\Theta \Theta}\, g_{\Phi \Phi}} \, d\Theta \,.
\end{equation}
The corresponding entropy $S_i=A_i/4$, $i=1,2$ of the extreme constituents is
\begin{eqnarray}\label{EK2}
S_1=S_2= (2-\sqrt{2}) \,\pi M^2\approx  0.5858  \times \pi M^2\,.
\end{eqnarray}
The total entropy of the system is 
\begin{eqnarray}\label{EK2}
S_{NEHK2}=S_1+S_2=(2-\sqrt{2}) \,2\pi M^2\,.
\end{eqnarray}

We find that the distance between the black holes decreases in the $M\rightarrow0$ limit. In this limit, the black holes in the NHEK2 solution (\ref{Weyl}) with functions (\ref{NHEK2sol2})-(\ref{NHEK2sol3}) collapse into one. The limiting metric is also finite, with \ref{Weyl} and functions
\begin{gather}\label{Weylalt}
f=\frac{4 M^2 \hat{\rho} ^2}{2 \hat{z}^2+\hat{\rho} ^2},\qquad \omega=\frac{\sqrt{\hat{z}^2+\hat{\rho} ^2}}{2 M^2} \,\\
 e^{2\nu}=\frac{M^2 \left(2 \hat{z}^2+\hat{\rho} ^2\right) }{\left(\hat{z}^2+\hat{\rho} ^2\right)^2}.
\end{gather}
The collapse geometry corresponds to the single NHEK black hole metric. Each of the black holes in the original NHEK2 geometry contributes a mass $M/2$. The larger NHEK black hole formed after the collapse has mass $M_{NHEK}=M$. The entropy in this case is
\begin{eqnarray}\label{EK}
S_{NHEK}= \,2\pi M^2\,.
\end{eqnarray}

Having the explicit expressions for the entropy of the NHEK2 geometries (\ref{EK2}) and NHEK (\ref{EK}), we can now check that indeed the black hole area theorem established by Hawking \cite{Hawking:1971tu} is satisfied for extremal black holes
\begin{eqnarray}
S_{NEHK2}&\le&S_{NHEK}\,.
\end{eqnarray}
A more detailed discussion about the entropy of extremal binary black holes (\ref{EK2}) and Hawking's area theorem can be found in \cite{Majo2018}.\\

{\bf Conical Singularity.} We now compute the conical singularities of the new NHEK2 metric 
\begin{equation}
\Delta\hat{\phi}= 2\pi \lim_{\hat{\rho}\rightarrow 0} \left(1-\sqrt{\frac{f}{\hat{\rho}^2 e^{2\nu}} }\right)\,, \quad -M<z<M \,,
\end{equation}
While the NHEK2 metric has no naked curvature singularities, our computation show that there is a non-removable conical deficit between the two black holes
\begin{equation}
\Delta\hat{\phi} =2\pi\,(1/2+\sqrt{2})\,.
%= -0.656854
\end{equation}

{\bf Charges and Entropy.}
The mass and angular momenta of the solution follow from the limiting near-horizon procedure of the extreme BBHs solution that is manifestly asymptotically flat
\begin{gather}
 M_1=M_2=M/2\,,\qquad J_1=J_2=M^2/2\,,
 \end{gather}
 Interestingly, the ratio $J_1/M_1^2=J_2/M_2^2=2$ is fixed. 
While Hawking's temperature is zero for NHEK2, each black hole has a non-trivial angular velocity
\begin{gather}
 \Omega_1=\Omega_2=1/ (2 M)\,,
 \end{gather}
 which displays its non-supersymmetric character.  It is straightforward to verify that the Smarr law (for the extremal, zero temperature configuration) is satisfied for each individual black hole $M_i= 2 \Omega_i J_i$.\\
 
{\bf Ergosphere.}
There are regions in the NHEK2 spacetime where the vector $\partial_{\hat{t}}$ becomes null. We will refer to the boundary region as the ergosphere, since they appear as a consequence of the presence of such regions in the original stationary extreme BBHs geometries. For NHEK2 these are defined by regions where $g^{\hat{t}\hat{t}}=0$ and give rise to a set of disconnected regions as shown in Fig. 1. The event horizons of the black holes in NHEK2 are points in the $(\rho,z)$-plane and have finite event horizon areas. There is a self similar behavior close to each black hole that resembles the ergospheres of isolated extremal Kerr black hole.

\begin{figure}[h!]
\begin{center}
\includegraphics[width=7.5cm,height=7.5cm]{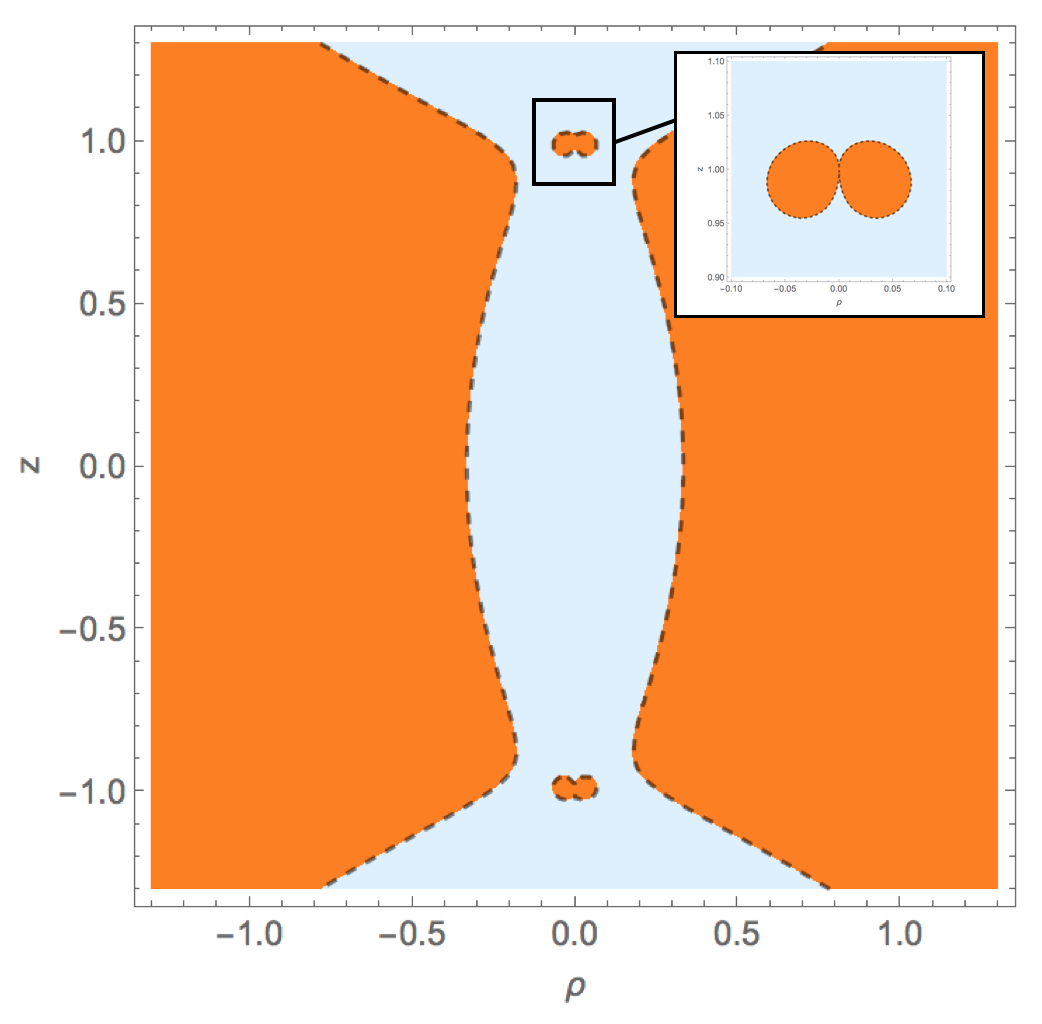}
\caption{\small Ergoregion (shaded {\it orange region}) of the NHEK2 black hole solution for $M=1$, as a consequence of the presence of such regions in the original stationary extreme BBHs geometries. The black holes are located at $\rho=\hat{\rho}_H=0, z=\hat{z}_H=\pm 1$. A more detailed diagram, close to one of the black holes appears in the upper right corner. The dashed line corresponds to the boundary where $\partial_{\hat{t}}$ is null.}
\label{ergo}
\end{center}
\end{figure}

{\bf Uniqueness.} The NHEK2 geometry, in contrast with NHEK, is not unique. For a fixed value of the total mass, it is expected that other finite near-horizon extremal BBHs exist. These may include extremal BBHs with spins that are not aligned. Another intriguing aspects related to the uniqueness includes the classifications of near horizon geometries in \cite{Kunduri:2007vf,Kunduri:2013ana}. NHEK2 does not pertain to any of the classes considered therein. This suggests that a new set of assumptions are needed for a complete  classification i.e.\ restricting the number of isolated horizons in the solutions.\\

{\bf Discussion.} The extremal BBHs solutions \cite{Manko:2011ts}, as well as the new NHEK2 geometries that we have found in this note, contain a localized conical singularity (strut), in the line separating the two black holes. Conical singularities are, in contrast to curvature singularities in the geometry, possibly avoidable as the spin orbit is reinstated. In the stationary BBHs, for example, one could introduce a rotation along one of the planes perpendicular to the angular momenta that are pointing along the azimuthal-axis. This mechanism would give rise to a centrifugal force, that may balance the self-gravitational effects between the two black holes and eliminate the strut. So far, however, this modification of the stationary BBHs solutions has not been achieved analytically. Other solutions where rotation plays a role in balancing different configurations have been given in \cite{Emparan:2001wk}-\cite{Lunin:2001fv}.
Another reason for considering the stationary rather than the dynamical BBHs solutions is their extremal maximally rotating (neutral) counterparts. Progress has been reported on rapidly spinning dynamical BBHs mergers \cite{Rezzolla:2007xa}-\cite{Lovelace:2010ne}, but extremal zero temperature BBHs are not viable yet through numerical explorations. Therefore, to study the extremal BBHs one would have to choose to work with the exact extreme stationary BBHs and consider the effect of the conical singularities in these solutions. While conical singularities are not desirable, these have been shown to be irrelevant to the so called black hole shadows \cite{Cunha:2018gql,Cunha} thanks to the underlying cylindrical symmetry of the problem. This suggests that certain physical properties of black holes are protected by symmetries and can be studied even with solutions that contain localized conical singularities. 

{\bf Acknowledgments.} The authors would like to thank  P. Cunha, S. Hadar and C. Herdeiro for helpful discussions and O. Varela for careful reading of the manuscript. This work was supported by the NSF grant PHY-1707571 at Utah State University and the Max Planck Gesellschaft through the Gravitation and Black Hole Theory Independent Research Group.

% Reference section using BibTeX:
%\bibliographystyle{plainnat}
%%%%%%%%%%%%%%%%%% BIBLIOGRAPHY %%%%%%%%%%%%%%%%%%%%


\begin{thebibliography}{99}


 %\cite{Gou:2011nq}
\bibitem{Gou:2011nq} 
  L.~Gou {\it et al.},
  ``The Extreme Spin of the Black Hole in Cygnus X-1,''
  Astrophys.\ J.\  {\bf 742}, 85 (2011)
 % doi:10.1088/0004-637X/742/2/85
  [arXiv:1106.3690 [astro-ph.HE]].
  %%CITATION = doi:10.1088/0004-637X/742/2/85;%%
  %121 citations counted in INSPIRE as of 09 Apr 2018
  
  
%\cite{Pounds:2003eb}
\bibitem{Pounds:2003eb} 
  K.~A.~Pounds, J.~N.~Reeves, A.~R.~King and K.~L.~Page,
  ``Exploring the complex x-ray spectrum of NGC 4051,''
  Mon.\ Not.\ Roy.\ Astron.\ Soc.\  {\bf 350}, 10 (2004)
%  doi:10.1111/j.1365-2966.2004.07639.x
  [astro-ph/0310257].
  %%CITATION = doi:10.1111/j.1365-2966.2004.07639.x;%%
  %72 citations counted in INSPIRE as of 10 Apr 2018

%\cite{Risaliti:2013cbe}
\bibitem{Risaliti:2013cbe} 
  G.~Risaliti {\it et al.},
  ``A rapidly spinning supermassive black hole at the centre of NGC 1365,''
  Nature {\bf 494}, 449 (2013)
%  doi:10.1038/nature11938
  [arXiv:1302.7002 [astro-ph.HE]].
  %%CITATION = doi:10.1038/nature11938;%%
  %127 citations counted in INSPIRE as of 10 Apr 2018
  
  %\cite{Fabian:2015xag}
\bibitem{Fabian:2015xag} 
  A.~C.~Fabian,
  ``The Innermost Extremes of Black Hole Accretion,''
  Astron.\ Nachr.\  {\bf 337}, no. 4/5, 375 (2017)
%  doi:10.1002/asna.201612316
  [arXiv:1511.08596 [astro-ph.HE]].
  %%CITATION = doi:10.1002/asna.201612316;%%
  %4 citations counted in INSPIRE as of 10 Apr 2018
  
 
    %\cite{McClintock:2006xd}
\bibitem{McClintock:2006xd} 
  J.~E.~McClintock, R.~Shafee, R.~Narayan, R.~A.~Remillard, S.~W.~Davis and L.~X.~Li,
  ``The Spin of the Near-Extreme Kerr Black Hole GRS 1915+105,''
  Astrophys.\ J.\  {\bf 652}, 518 (2006)
%  doi:10.1086/508457
  [astro-ph/0606076].
  %%CITATION = doi:10.1086/508457;%%
  %326 citations counted in INSPIRE as of 09 Apr 2018 
 

%\cite{Rezzolla:2007xa}
\bibitem{Rezzolla:2007xa} 
  L.~Rezzolla, E.~N.~Dorband, C.~Reisswig, P.~Diener, D.~Pollney, E.~Schnetter and B.~Szilagyi,
  ``Spin Diagrams for Equal-Mass Black-Hole Binaries with Aligned Spins,''
  Astrophys.\ J.\  {\bf 679}, 1422 (2008)
%  doi:10.1086/587679
  [arXiv:0708.3999 [gr-qc]].
  %%CITATION = doi:10.1086/587679;%%
  %88 citations counted in INSPIRE as of 10 Apr 2018

%\cite{Kesden:2009ds}
\bibitem{Kesden:2009ds} 
  M.~Kesden, G.~Lockhart and E.~S.~Phinney,
  ``Maximum black-hole spin from quasi-circular binary mergers,''
  Phys.\ Rev.\ D {\bf 82}, 124045 (2010)
%  doi:10.1103/PhysRevD.82.124045
  [arXiv:1005.0627 [gr-qc]].
  %%CITATION = doi:10.1103/PhysRevD.82.124045;%%
  %23 citations counted in INSPIRE as of 10 Apr 2018
  
  %\cite{Lovelace:2010ne}
\bibitem{Lovelace:2010ne} 
  G.~Lovelace, M.~A.~Scheel and B.~Szilagyi,
  ``Simulating merging binary black holes with nearly extremal spins,''
  Phys.\ Rev.\ D {\bf 83}, 024010 (2011)
%  doi:10.1103/PhysRevD.83.024010
  [arXiv:1010.2777 [gr-qc]].
  %%CITATION = doi:10.1103/PhysRevD.83.024010;%%
  %71 citations counted in INSPIRE as of 09 Apr 2018

  
%\cite{Volonteri:2004cf}
\bibitem{Volonteri:2004cf} 
  M.~Volonteri, P.~Madau, E.~Quataert and M.~J.~Rees,
  ``The Distribution and cosmic evolution of massive black hole spins,''
  Astrophys.\ J.\  {\bf 620}, 69 (2005)
 % doi:10.1086/426858
  [astro-ph/0410342].
  %%CITATION = doi:10.1086/426858;%%
  %207 citations counted in INSPIRE as of 10 Apr 2018
  
%\cite{Berti:2008af}
\bibitem{Berti:2008af} 
  E.~Berti and M.~Volonteri,
  ``Cosmological black hole spin evolution by mergers and accretion,''
  Astrophys.\ J.\  {\bf 684}, 822 (2008)
  %doi:10.1086/590379
  [arXiv:0802.0025 [astro-ph]].
  %%CITATION = doi:10.1086/590379;%%
  %197 citations counted in INSPIRE as of 10 Apr 2018

[6] K. S. Thorne, Astrophys. J. 191, 507 (1974).
%\cite{Gammie:2003qi}
\bibitem{Gammie:2003qi} 
  C.~F.~Gammie, S.~L.~Shapiro and J.~C.~McKinney,
  ``Black hole spin evolution,''
  Astrophys.\ J.\  {\bf 602}, 312 (2004)
%  doi:10.1086/380996
  [astro-ph/0310886].
  %%CITATION = doi:10.1086/380996;%%
  %199 citations counted in INSPIRE as of 10 Apr 2018
  
%\cite{Shapiro:2004ud}
\bibitem{Shapiro:2004ud} 
  S.~L.~Shapiro,
  ``Spin, accretion and the cosmological growth of supermassive black holes,''
  Astrophys.\ J.\  {\bf 620}, 59 (2005)
%  doi:10.1086/427065
  [astro-ph/0411156].
  %%CITATION = doi:10.1086/427065;%%
  %135 citations counted in INSPIRE as of 10 Apr 2018
  
%\cite{Bardeen:1999px}
\bibitem{Bardeen:1999px} 
  J.~M.~Bardeen and G.~T.~Horowitz,
  ``The Extreme Kerr throat geometry: A Vacuum analog of AdS(2) x S**2,''
  Phys.\ Rev.\ D {\bf 60}, 104030 (1999)
%  doi:10.1103/PhysRevD.60.104030
  [hep-th/9905099].
  %%CITATION = doi:10.1103/PhysRevD.60.104030;%%
  %291 citations counted in INSPIRE as of 10 Apr 2018

%\cite{Guica:2008mu}
\bibitem{Guica:2008mu} 
  M.~Guica, T.~Hartman, W.~Song and A.~Strominger,
  ``The Kerr/CFT Correspondence,''
  Phys.\ Rev.\ D {\bf 80}, 124008 (2009)
%  doi:10.1103/PhysRevD.80.124008
  [arXiv:0809.4266 [hep-th]].
  %%CITATION = doi:10.1103/PhysRevD.80.124008;%%
  %515 citations counted in INSPIRE as of 10 Apr 2018
  
 %\cite{Hadar:2014dpa}
\bibitem{Hadar:2014dpa} 
  S.~Hadar, A.~P.~Porfyriadis and A.~Strominger,
  ``Gravity Waves from Extreme-Mass-Ratio Plunges into Kerr Black Holes,''
  Phys.\ Rev.\ D {\bf 90}, no. 6, 064045 (2014)
 % doi:10.1103/PhysRevD.90.064045
  [arXiv:1403.2797 [hep-th]].
  %%CITATION = doi:10.1103/PhysRevD.90.064045;%%
  %29 citations counted in INSPIRE as of 10 Apr 2018
  
    %\cite{Lupsasca:2014pfa}
\bibitem{Lupsasca:2014pfa} 
  A.~Lupsasca, M.~J.~Rodriguez and A.~Strominger,
  ``Force-Free Electrodynamics around Extreme Kerr Black Holes,''
  JHEP {\bf 1412}, 185 (2014)
%  doi:10.1007/JHEP12(2014)185
  [arXiv:1406.4133 [hep-th]].
  %%CITATION = doi:10.1007/JHEP12(2014)185;%%
  %23 citations counted in INSPIRE as of 10 Apr 2018
  
  %\cite{Kinnersley:1978pz}
\bibitem{Kinnersley:1978pz} 
  W.~Kinnersley and D.~M.~Chitre,
  ``Symmetries of the stationary Einstein–Maxwell equations. IV. Transformations which preserve asymptotic flatness,''
  J.\ Math.\ Phys.\  {\bf 19}, 2037 (1978).
% doi:10.1063/1.523580
  %%CITATION = doi:10.1063/1.523580;%%  
  
%\cite{Manko:2009zz}
\bibitem{Manko:2009zz} 
  V.~S.~Manko, E.~D.~Rodchenko, E.~Ruiz and B.~I.~Sadovnikov,
  ``On the simplest binary system of rotating black holes,''
  AIP Conf.\ Proc.\  {\bf 1122}, 332 (2009).
%  doi:10.1063/1.3141316
  %%CITATION = doi:10.1063/1.3141316;%%
  
     %\cite{Herdeiro:2008kq}
\bibitem{Herdeiro:2008kq} 
  C.~A.~R.~Herdeiro and C.~Rebelo,
  ``On the interaction between two Kerr black holes,''
  JHEP {\bf 0810}, 017 (2008)
  [arXiv:0808.3941 [gr-qc]].
  %%CITATION = ARXIV:0808.3941;%%
  %17 citations counted in INSPIRE as of 11 mar 2015
  
  
%\cite{Manko:2008pv}
\bibitem{Manko:2008pv} 
  V.~S.~Manko, E.~D.~Rodchenko, E.~Ruiz and B.~I.~Sadovnikov,
  ``Exact solutions for a system of two counter-rotating black holes,''
  Phys.\ Rev.\ D {\bf 78}, 124014 (2008)
%  doi:10.1103/PhysRevD.78.124014
  [arXiv:0809.2422 [gr-qc]].
  %%CITATION = doi:10.1103/PhysRevD.78.124014;%%
  %9 citations counted in INSPIRE as of 10 Apr 2018
  
  \bibitem{BZ}   
  V.~A.~Belinski nd V.~E.~ Zakharov, 
  ``Stationary gravitational solitons with axial symmetry,"
  Sov.\  Phys.\ JETP {\bf50(1)} 1 (1979)
  
   \bibitem{Kramer} 
  D.~Kramer and G.~Neugebauer, 
  ''The superposition of two Kerr solutions," 
  Phys.\ Lett.\ A. {\bf 75 4} 259 (1980).
  
%\cite{Costa:2009wj}
\bibitem{Costa:2009wj} 
  M.~S.~Costa, C.~A.~R.~Herdeiro and C.~Rebelo,
  ``Dynamical and Thermodynamical Aspects of Interacting Kerr Black Holes,''
  Phys.\ Rev.\ D {\bf 79}, 123508 (2009)
  [arXiv:0903.0264 [gr-qc]].
  %%CITATION = ARXIV:0903.0264;%%
  %11 citations counted in INSPIRE as of 23 Feb 2015
  

  %\cite{Manko:2011ts}
\bibitem{Manko:2011ts} 
  V.~S.~Manko and E.~Ruiz,
  ``On a Simple Representation of the Kinnersley-Chitre Metric,''
  Prog.\ Theor.\ Phys.\  {\bf 125}, 1241 (2011)
  [arXiv:1101.4687 [gr-qc]].
  %%CITATION = ARXIV:1101.4687;%%
  %1 citations counted in INSPIRE as of 23 Feb 2015

%\cite{Maldacena:1998uz}
\bibitem{Maldacena:1998uz} 
  J.~M.~Maldacena, J.~Michelson and A.~Strominger,
  ``Anti-de Sitter fragmentation,''
  JHEP {\bf 9902}, 011 (1999)
  [hep-th/9812073].
  %%CITATION = HEP-TH/9812073;%%
  %230 citations counted in INSPIRE as of 23 Feb 2015
  
 \bibitem{Hawking:1971tu} 
  S.~W.~Hawking,
  ``Gravitational radiation from colliding black holes,''
  Phys.\ Rev.\ Lett.\  {\bf 26}, 1344 (1971).
%  doi:10.1103/PhysRevLett.26.1344
  %%CITATION = doi:10.1103/PhysRevLett.26.1344;%%
  %600 citations counted in INSPIRE as of 31 Mar 2018
 
\bibitem{Majo2018} 
    M.~J.~Rodriguez,
  ``Extremal Binary Black Hole Entropy,'' {\it to appear}.
  
    %\cite{Kunduri:2007vf}
\bibitem{Kunduri:2007vf} 
  H.~K.~Kunduri, J.~Lucietti and H.~S.~Reall,
  ``Near-horizon symmetries of extremal black holes,''
  Class.\ Quant.\ Grav.\  {\bf 24}, 4169 (2007)
%  doi:10.1088/0264-9381/24/16/012
  [arXiv:0705.4214 [hep-th]].
  %%CITATION = doi:10.1088/0264-9381/24/16/012;%%
  %221 citations counted in INSPIRE as of 16 Apr 2018
  
  %\cite{Kunduri:2013ana}
\bibitem{Kunduri:2013ana} 
  H.~K.~Kunduri and J.~Lucietti,
  ``Classification of near-horizon geometries of extremal black holes,''
  Living Rev.\ Rel.\  {\bf 16}, 8 (2013)
%  doi:10.12942/lrr-2013-8
  [arXiv:1306.2517 [hep-th]].
  %%CITATION = doi:10.12942/lrr-2013-8;%%
  %92 citations counted in INSPIRE as of 16 Apr 2018
  
  %\cite{Emparan:2001wk}
\bibitem{Emparan:2001wk} 
  R.~Emparan and H.~S.~Reall,
  ``Generalized Weyl solutions,''
  Phys.\ Rev.\ D {\bf 65}, 084025 (2002)
%  doi:10.1103/PhysRevD.65.084025
  [hep-th/0110258].
  %%CITATION = doi:10.1103/PhysRevD.65.084025;%%
  %232 citations counted in INSPIRE as of 31 Mar 2018
  
  %\cite{Emparan:2001ux}
\bibitem{Emparan:2001ux} 
  R.~Emparan, D.~Mateos and P.~K.~Townsend,
  ``Supergravity supertubes,''
  JHEP {\bf 0107}, 011 (2001)
%  doi:10.1088/1126-6708/2001/07/011
  [hep-th/0106012].
  %%CITATION = doi:10.1088/1126-6708/2001/07/011;%%
  %162 citations counted in INSPIRE as of 31 Mar 2018
  
    
  %\cite{Lunin:2001fv}
\bibitem{Lunin:2001fv} 
  O.~Lunin and S.~D.~Mathur,
  ``Metric of the multiply wound rotating string,''
  Nucl.\ Phys.\ B {\bf 610}, 49 (2001)
%  doi:10.1016/S0550-3213(01)00321-2
  [hep-th/0105136].
  %%CITATION = doi:10.1016/S0550-3213(01)00321-2;%%
  %173 citations counted in INSPIRE as of 31 Mar 2018
 
 %\cite{Cunha:2018gql}
\bibitem{Cunha:2018gql} 
  P.~V.~P.~Cunha, C.~A.~R.~Herdeiro and M.~J.~Rodriguez,
  ``Does the black hole shadow probe the event horizon geometry?,''
  arXiv:1802.02675 [gr-qc].
  %%CITATION = ARXIV:1802.02675;%%
  %1 citations counted in INSPIRE as of 10 Apr 2018
  
   %\cite{Cunha}
\bibitem{Cunha}
   P.~V.~P.~Cunha, C.~A.~R.~Herdeiro and M.~J.~Rodriguez,
  ``Shadows of Exact Binary Black Holes,'' {\it to appear}.
    
  \end{thebibliography}
\end{document}